\documentclass[prl,twocolumn]{revtex4-1}

\usepackage{amsmath}
\usepackage{amssymb}
\usepackage{bm}
\usepackage{color}
\usepackage{graphicx}
\usepackage[latin1]{inputenc}
\usepackage[english]{babel}
\usepackage{sistyle}

\begin{document}

\title{Experimental evidence of self-localized and propagating spin wave modes in obliquely magnetized current-driven nanocontacts}

\author{Stefano Bonetti$^1$}
\email{bonetti@kth.se}
\author{Vasil Tiberkevich$^{2}$}
\author{Giancarlo Consolo$^{3,4}$}
\author{Giovanni Finocchio$^{4}$}
\author{Pranaba Muduli$^{5}$}
\author{Fred Mancoff$^{6}$}
\author{Andrei Slavin$^{2}$}
\author{Johan \AA kerman$^{1,5}$}

\affiliation{
$^1$~Materials Physics, Royal Institute of Technology, Electrum 229, 164 40 Kista, Sweden
\\
$^2$~Department of Physics, Oakland University, Rochester, MI 48309 USA
\\
$^3$~Department of Physics, University of Ferrara, Ferrara, Italy
\\
$^4$~Department of Matter Physics and Electronic Engineering, University of Messina, Messina, Italy
\\
$^5$~Department of Physics, University of Gothenburg, Gothenburg, Sweden
\\
$^6$~Everspin Technologies, Inc., 1300 N. Alma School Road, Chandler AZ, USA
}

%
%
%
%
%
%
%
%

\begin{abstract}

Through detailed experimental studies of the angular dependence of spin wave excitations in nanocontact-based spin-torque oscillators, we demonstrate that two distinct spin wave modes can be excited, with different frequency, threshold currents and frequency tuneability. Using analytical theory and micromagnetic simulations we identify one mode as an exchange-dominated propagating spin wave, and the other as a self-localized nonlinear spin wave bullet. Wavelet-based analysis of the simulations indicates that the apparent simultaneous excitation of both modes results from rapid mode hopping induced by the Oersted field. 

\end{abstract}

\maketitle

Spin-polarized currents passing through a thin magnetic film can excite spin waves via the spin-transfer-torque effect \cite{Slonczewski1996,Berger1996}. In his pioneering paper \cite{Slonczewski1999}, Slonczewski predicted that such spin waves, excited in perpendicularly magnetized free layer underneath a nanocontact \cite{PhysRevLett.80.4281}, would be exchange-dominated, propagating radially from the nanocontact region, with a wave number $k$ inversely proportional to the nanocontact radius $R_c$ ($k \simeq 1.2/R_c$). Rippard \emph{et al.} \cite{Rippard2003APL} subsequently demonstrated that, while Slonczewski's theory correctly describes the frequency and threshold current of spin waves excited in perpendicularly magnetized films, it fails to describe spin waves excited when the same film is magnetized in the plane. It was later shown theoretically that in the case of in-plane magnetization, apart from the Slonczewski-like propagating spin wave mode, it is possible to excite a self-localized nonlinear spin wave mode of solitonic character -- a so-called standing spin wave bullet \cite{Slavin2005PRL}. While the current-induced excitation of the spin wave bullet was subsequently confirmed in several numerical simulations \cite{consolo:144410, berkov:144414, consolo:014420, Berkov20081238}, and angular dependent measurements have been presented in the literature \cite{Rippard2004a}, no clear experimental evidence of a spin wave bullet nor a characterization of its properties, has yet been presented.

In this Letter, we study the angular dependence of spin wave excitations in nanocontact based spin-torque oscillators (STO), and demonstrate that when the free layer is magnetized at sufficiently small angles $\theta_e \lesssim 55^\circ$ w.r.t the plane, two distinct and qualitatively different spin wave modes can be excited by the current passing through the nanocontact. The two modes have different frequencies, $f$, different threshold currents, $I_{th}$, and opposite sign of the frequency tuneability, $df/dI$. Through comparison with theory \cite{Slavin2005PRL,gerhart:024437} and micromagnetic simulations, we show that the higher-frequency, blue-shifting mode is well described by the Slonczewski propagating mode, and that the lower-frequency, red-shifting mode exhibits all the predicted properties of a localized spin wave bullet. Using time-frequency wavelet-based analysis of our micromagnetic simulations we furthermore demonstrate that the apparent simultaneous excitation of both modes, as observed in our frequency domain experiments, results from non-stationary switching between the two modes on the sub-ns time scale.

The magnetically active part of the sample is a Co$_{81}$Fe$_{19}$(20~nm)/Cu(6~nm)/Ni$_{80}$Fe$_{20}$(4.5~nm) thin
film tri-layer, patterned into a 8~$\mu$m~$\times$~26~$\mu$m mesa.
On top of this mesa, a circular Al nanocontact having nominal diameter $2R_c = 40$~nm was defined through SiO$_2$ using e-beam lithography (see Ref. \cite{Mancoff2006} for details). An external magnetic field of constant magnitude ($\mu_0H_e=1.1$ T) was applied to the sample at an angle $\theta_e$ w.r.t the film plane. Details of the measurements setup are given in Ref.~\cite{bonetti:102507}. The excited spin waves modulate the magnetoresistance of the device and are detected as a microwave voltage signal. Microwave excitations were only observed for a single current polarity, corresponding to electrons flowing from the ``free'' (thin NiFe) to the ``fixed'' (thick CoFe) magnetic layer. All measurements were performed at room temperature. While the results presented here all come from a single sample, we have confirmed that the results obtained on several other devices are qualitatively similar.

\begin{figure}[t]
\centering
\includegraphics[width=0.48\textwidth]{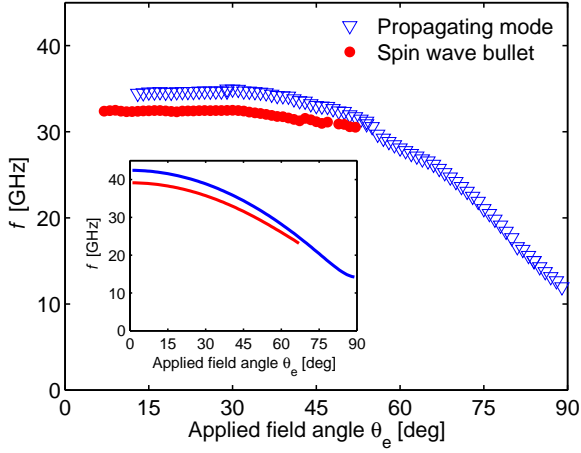}
\caption{(Color online) Measured frequencies of the observed spin wave modes as a function of the applied field angle $\theta_e$ at $I = 14$~mA and $\mu_0H_e = 1.1$~T. Inset: theoretically calculated frequencies of the propagating (upper curve) and the bullet (lower curve) modes at the current threshold, for nominal parameters of the nanocontact STO.}
\label{fig:f_vs_theta}
\end{figure}

Figure~\ref{fig:f_vs_theta} shows the angular dependence of the microwave frequencies generated at a constant current of $I = 14$~mA.
The most striking feature is the existence of two distinct signals for sufficiently small values of $\theta_e$. The frequencies of these two signals differ by about 2.5 GHz at angles up to $\theta_e = 40^\circ$, and then start to approach each other up to a critical angle $\theta_e = \theta_c \approx 55^\circ$ where the lower frequency signal disappears. The general behavior of the two signals remains the same at higher currents ($I = 18$~mA), with a slight increase of both $\theta_c$ and the frequency separation to about $58^\circ$ and 3 GHz respectively. 

\begin{figure}[t]
\includegraphics[width=0.48\textwidth]{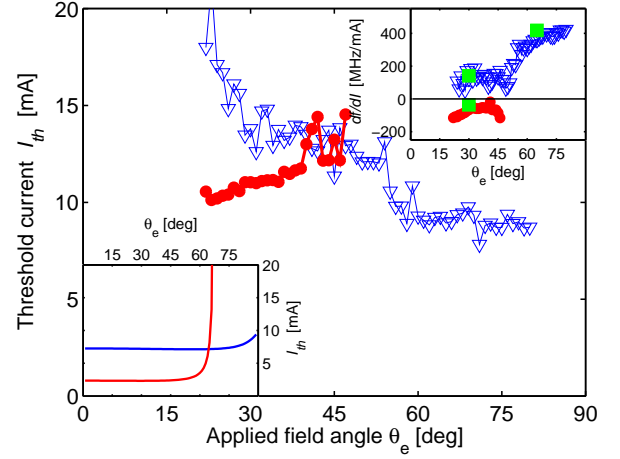}
\caption{(Color online) Measured threshold current for the propagating (empty triangles) and the bullet (filled circles) modes as a function of applied field angle $\theta_e$. Lower inset: theoretical threshold current vs applied field angle. Upper inset: $df/dI$ vs. $\theta_e$ for the propagating (empty triangles) and the bullet (filled circles) modes as a function of $\theta_e$. Filled squares are the results of micromagnetic simulations.} \label{fig:Ith_vs_theta}
\end{figure}

Fig.~\ref{fig:Ith_vs_theta} shows the angular dependence of the threshold current $I_{th}$ for both signals. $I_{th}$ is found from peak power ($p$) measurements in the sub-critical regime ($I<I_{th}$) analyzed using the method proposed in Refs.~\cite{tiberkevich:192506, SlavinTutorial}, and employed in Ref.~\cite{kudo:07D105}. Since $1/p(I) \sim
(I_{th}-I)$ in this regime, $I_{th}$ for each signal can be directly determined from the intercept of a
straight line through $1/p(I)$ vs. $I$ with the current axis. We extracted $I_{th}$ only for the magnetization angles $20^\circ <
\theta_e < 80^\circ$, since outside this range the signal was too
weak to allow for a reliable analysis. We note that the low-frequency signal always has the lower threshold
current (within the noise of the analysis), in particular at low field angles. As the angle increases, the $I_{th}$ values for the two signals gradually approach each other and become
essentially equal close to $\theta_c$. For the lower-frequency signal, the data is plotted up to $\theta_e=47^\circ$, since above this angle the signal is too low to allow for a reliable determination of $I_{th}$. 

\begin{figure}[t]
\centering
\includegraphics[width=0.50\textwidth]{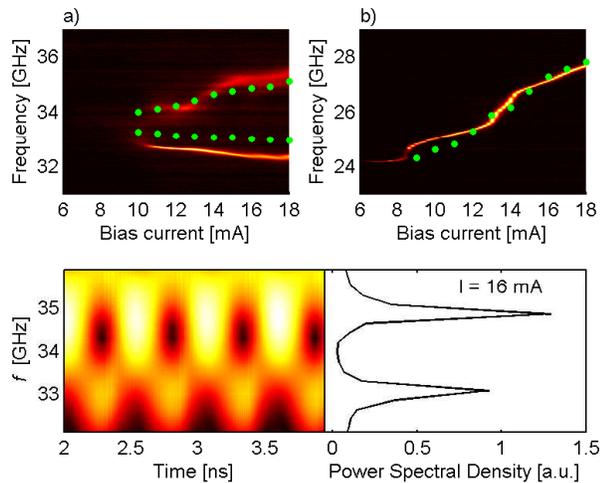}
\caption{(Color online) Comparison of experiment with micromagnetic simulations: (a, b) Measured microwave power, presented as color maps onto the frequency-current plane, for two applied field angles (a) $\theta_e = 30^\circ$ and (b) $\theta_e = 65^\circ$, where filled symbols show the results of the micromagnetic simulations. (c) Wavelet analysis, and (d) Fast Fourier transform, of the micromagnetic simulation at $\theta_e = 30^\circ$ and $I=16$ mA.}
\label{fig:map_and_threshold}
\end{figure}

The upper inset in Fig.~\ref{fig:Ith_vs_theta} shows the
experimental tuneabilities $df/dI$ of the two signals. The lower-frequency signal always red shifts with current ($df/dI<0$) with values ranging from -40 to -110 MHz/mA. In contrast,
the higher-frequency signal blue shifts with current ($df/dI>0$) with values ranging from +50 to +150 MHz/mA for $\theta_e < 45^\circ$ and from +300 to +400 MHz/mA when $\theta_e > 55^\circ$. The opposite tuneability sign can be clearly seen in Fig.~\ref{fig:map_and_threshold}a-b where the microwave power of both signals is color mapped onto the frequency-current plane.
At $\theta_e = 30^\circ$ (Fig.~\ref{fig:map_and_threshold}a)) both signals are visible with a lower $I_{th}$ and a clear red shift for the lower-frequency signal, and a higher $I_{th}$ and a clear blue shift for the higher-frequency signal. At a larger magnetization angle, $\theta_e = 65^\circ$ (Fig.~\ref{fig:map_and_threshold}b)), only the higher-frequency, blue-shifting signal is visible.

In order to gain a physical understanding of the experimental observations, we use the theoretical model of spin wave excitations developed in Refs.~\cite{Slavin2005PRL, gerhart:024437}. In this model, one of the excited modes is directly related to Slonczewski's propagating spin wave mode for a perpendicularly magnetized nanocontact STO \cite{Slonczewski1999}, but now generalized for the case of an oblique orientation of the free layer magnetization. The mode retains its propagating character,  i.e. it carries energy away from the nanocontact area, and its threshold current is approximately given by:
\begin{equation}\label{th-linear}
    I_{th}^{prop} \approx \left [\Gamma(\theta_e)+ 1.86 D(\theta_e)/R_c^2 \right ]/\sigma(\theta_e)
\ ,\end{equation}where $\Gamma(\theta_e)$ is the Gilbert damping rate in the free layer,
$D(\theta_e)$ is the spin wave dispersion coefficient, $R_c$ is the nano-contact
radius, and $\sigma(\theta_e)$ is the spin-polarization pre-factor (see Eq.~(2) in Ref. \cite{Slavin2005PRL}). The second term in Eq. (\ref{th-linear}) is independent of the spin wave
damping $\Gamma$ and describes radiative energy losses due to the
propagating character of the mode. It is usually larger than the
first term, which describes losses due to direct energy dissipation
in the nanocontact area. The frequency of the propagating mode
\begin{equation}\label{freq-linear}
    \omega_{prop} \approx \omega_0(\theta_e) +
    1.44D(\theta_e)/R_c^2+N(\theta_e)|a|^2
\end{equation}
is hence typically {\em higher} than the FMR frequency $\omega_0$ of the free layer, and depending on the sign of the nonlinear frequency shift, $N=\partial\omega/\partial|a_0|^2 $, either blue shifts or red shifts with increasing current.

The model also predicts the existence of an additional mode whose amplitude 
and spatial extension are stabilized by two concurring processes:
\emph{i}) energy dissipation due to Gilbert damping and \emph{ii})
energy gain from the spin-polarized current. This mode is self-localized, has a two-dimensional solitonic character with a bullet shaped amplitude profile (hence also called a spin wave bullet), and does not carry any energy away from the
nanocontact area. The expression for its threshold current consequently lacks the radiative term, and, therefore, $I_{th}^{bul}$ is  directly
proportional to the spin wave damping:
\begin{equation}\label{th-bullet}
    I_{th}^{bul} \approx \beta\Gamma(\theta_e)/\sigma(\theta_e)
\,,\end{equation} where the dimensionless coefficient $\beta \sim 1$
depends on the parameters of the system (for further details see
\cite{Slavin2005PRL}). As a consequence, the spin wave bullet mode
always has a \emph{lower} threshold current than the
Slonczewski-like propagating mode, except very close to $\theta_{cr}$ where $\beta$ diverges.

The spin wave bullet mode does not exist at all magnetization angles. Only for $N<0$ can the nonlinearity counteract the dispersion-related spreading of the spin wave profile. Since $N$ is a function of the magnetization angle, and changes sign from always negative for an in-plane magnetization to always positive for a perpendicular magnetization \cite{SlavinTutorial}, the spin wave bullet mode only exists for applied field angles smaller than a certain critical value $\theta_{cr}$, and does not exist
for $\theta_e > \theta_{cr}$ \cite{gerhart:024437}. As a direct consequence of the negative N, the frequency of the spin wave bullet mode 
\begin{equation}\label{freq-bullet}
    \omega_{bul} \approx \omega_0(\theta_e) + N(\theta_e)|a_0|^2
\end{equation}
always lies {\em below} the FMR frequency $\omega_0$, and continues to \emph{decrease} with increasing current (it always red shifts).

The inset of Fig.~\ref{fig:f_vs_theta} shows the calculated $f$ vs. $\theta_{e}$ for the two modes, using the nominal parameters of the measured STO. The experimental data confirms the theoretical predictions, such as the existence of a critical angle below which a lower-frequency mode is excited, and a qualitatively similar angular dependence of the frequency. The inset of Fig.~\ref{fig:Ith_vs_theta} shows $I_{th}$ vs.~$\theta_{e}$ for both modes. Again, the experimental data exhibits the same qualitative behavior as predicted by theory. In particular, $I_{th}$ of the lower-frequency mode is always lower than that of the higher-frequency mode. The combined qualitative agreement provides a strong argument for identifying the observed higher-frequency mode as a propagating mode, and the lower-frequency mode as a spin wave bullet.

There are, however, some notable differences between theory and our experiments, the most striking being the apparent \emph{simultaneous} excitation of both modes. Such a co-excitation is neither supported by theory \cite{Slavin2005PRL, gerhart:024437}, nor by micromagnetic simulations \cite{consolo:144410, consolo:014420}, where, on the contrary, a hysteresis between the two modes was observed. Another significant difference is the lack of any observed red shift of the propagating mode. While our experiments clearly demonstrate that the propagating mode is blue shifted at all field angles, Eq. (\ref{freq-linear}) predicts a red shift also for this mode when $N<0$, i.e.\ for $\theta_e < \theta_{cr}$, in clear contradiction with our experimental data. As we show below, both effects can be explained by taking into account the large Oersted field generated in the nanocontact, which was previously ignored in \cite{Slavin2005PRL, gerhart:024437, consolo:144410, consolo:014420}, but can be accounted for in micromagnetic simulations.

Micromagnetic simulations were carried out in a box shaped free layer volume $800\times800\times4.5$ nm$^3$ with constant cell size $4\times4\times4.5$ nm$^3$. A uniform spin polarized current acted on a quasi-cylindrical sub-volume of the free layer with an adjustable radius $R_c$. The material properties of the NiFe free layer were: saturation magnetization $\mu_0M_{S, \rm{free}}=0.7$ T,  Gilbert damping constant $\alpha_G=0.01$, exchange constant $A = 1.1\times10^{-11}$ J/m, and spin-torque efficiency $\epsilon=0.3$. The thickness and saturation magnetization of the CoFe fixed layer were 20 nm and $\mu_0M_{S,\rm{fixed}}=1.8$ T, respectively, and a minimization of the fixed layer magnetostatic energy in the applied field determined the fixed layer magnetization angle and consequently the polarization angle of the spin polarized current. No magneto-crystalline anisotropy, RKKY interaction, or dipolar coupling between the two magnetic layers were taken into account. The external field magnitude was fixed at $\mu_0H_{ext}=1.15$ T, and its direction was varied to fit the experimental data. All simulations were done at $T=0$ K.

The results of our micromagnetic simulations are shown as symbols in Fig.~\ref{fig:map_and_threshold}a-b after optimization of $R_c=32$ nm and magnetic field angles of 35$^\circ$ and 70$^\circ$ respectively. The quantitative agreement with the experimental data is remarkable: excitation frequency, range of the mode existence, and frequency tuneability of the two modes are well reproduced by the simulations. The reason for the larger simulated contact radius (32 nm vs. nominally 20 nm in the experiment) is likely due to two neglected effects: \emph{i}) current crowding at the contact perimeter, essentially increasing the effective contact radius compared to a uniform current, and \emph{ii}) lateral current spread in the free and fixed layers. The simulated field angles agrees with those used in the experiments to within 10\%. 

It is clear from our simulations that the inclusion of the Oersted field affects the magnetization dynamics in several significant ways.
First, the Oersted field has a strong qualitative impact on $df/dI$ of the propagating mode, making it positive at all investigated field angles, effectively resolving the apparent discrepancy between our experimental observations and theory. $df/dI$ of the bullet mode, on the other hand, remains largely unchanged with the inclusion of the Oersted field. If we now add the simulated $df/dI$ at 35$^\circ$ and 70$^\circ$ to the upper inset of Fig.~\ref{fig:Ith_vs_theta}, we also find a remarkable \emph{quantitative} agreement with the experimental values for both modes. 

Second, the inclusion of the Oersted field also makes the simulations reproduce the apparent simultaneous excitation of both modes. While our experimental setup is limited to measurements in the frequency domain, our micromagnetic simulations allow us to also investigate the detailed temporal evolution of the instantaneous power in each mode. Fig.~\ref{fig:map_and_threshold}c shows the result of a wavelet analysis \cite{PhysRevB.79.104438} of the micromagnetic data, where the dashed line indicates the instantaneous maximum power as a function of time. The analysis makes it clear that the two modes are in fact never excited at the same time. The instantaneous microwave power instead exhibits a persistent (at $T=0$ K also periodic) hopping between the two modes with a very high hopping frequency exceeding 1.5 GHz. It is noteworthy that we only observe such hopping when the Oersted field is properly included, which is likely related to the strong spatial inhomogeneities it induces in the vicinity of the nanocontact ~\cite{Consolo2007JAP, consoloIEEE2009}.

Finally, the influence of the Oersted field also explains the large quantitative discrepancy between the analytically calculated $I_{th}$ and the experimental value (Fig.~\ref{fig:Ith_vs_theta}). Previous micromagnetic simulations have demonstrated that the Oersted field  can cause a substantial (up to four-fold) increase of $I_{th}$ (see Fig.~2 in Ref.~\cite{consoloIEEE2009}), which hence agrees much better with our experimental data.

In conclusion, we have presented a detailed experimental study of the field angle dependence of spin wave excitations in nanocontact based spin-torque oscillators. We find that two distinct and qualitatively very different spin wave modes can be excited for applied field angles $\theta_e\lesssim55^\circ$. Through a comparison of our experimental measurements of three different fundamental properties ($f$, $df/dI$, and $I_{th}$) with both previously developed analytic theories \cite{Slonczewski1999,Slavin2005PRL,gerhart:024437} and our own micromagnetic simulations, we unambiguously identify the higher-frequency mode as an exchange-dominated propagating spin wave, and the lower-frequency mode as a self-localized non-propagating solitonic mode, i.e. a spin wave bullet. Our micromagnetic simulations show that not only is the Oersted field required to explain the sign of $df/dI$, and the magnitude of $I_{th}$, it is also solely responsible for the rapid (sub-ns) hopping between the two modes, which in frequency domain experiments make them appear as simultaneously excited.

We gratefully acknowledge financial support from The Swedish Foundation for strategic Research (SSF), the Swedish Research Council (VR), the G\"oran Gustafsson Foundation, the Knut and Alice Wallenberg Foundation, by the Contract no. W56HZV-09-P-L564 from the U.S. Army TARDEC and RDECOM, by the Grant no. ECCS-0653901 from the National Science Foundation of the USA, and by the Oakland University Foundation. Johan {\AA}kerman is a Royal Swedish Academy of Sciences Research Fellow supported by a grant from the Knut and Alice Wallenberg Foundation.

\end{document}